\documentclass[12pt]{article}
\let\Large=\large
\let\large=\normalsize

\newcommand{\be}[3]{\begin{equation}  \label{#1#2#3}}     
\newcommand{\bib}[3]{\bibitem{#1#2#3}}
\newcommand{\ee}{ \end{equation}}
\newcommand{\ba}{\begin{array}}
\newcommand{\ea}{\end{array}}

\begin{document}

\thispagestyle{empty}
\rightline{HUB-EP-97/81}
\rightline{hep-th/9710228}

\vspace{1.5truecm}

\centerline{\bf \Large On central charges and the entropy in matrix theory}
\vspace{2truecm}

\centerline{{\bf Klaus Behrndt}\footnote{e-mail: 
behrndt@qft2.physik.hu-berlin.de} 
 }
\vspace{.5truecm}
\centerline{\em Humboldt University Berlin}
\centerline{\em Invalidenstrasse 110, 10115 Berlin, Germany}

\vspace{2.2truecm}


\vspace{.5truecm}

\begin{abstract}

\noindent
The Bekenstein-Hawking entropy of BPS black holes can be obtained as
the minimum of the mass (= largest central charge).  In this letter we
investigate the analog procedure for the matrix model of $M$-theory.
Especially we discuss the configurations: (i) $2 \times 2 \times 2$
corresponding to the 5-d black hole and (ii) the $5 \times 5 \times 5$
configuration yielding the 5-d string. After getting their
matrix-entropy, we discuss a way of counting of microstates in matrix
theory. As Yang Mills field theory we propose the gauged world volume
theory of the 11-d KK monopole.

\end{abstract}

\bigskip \bigskip
\newpage

\noindent
{\bf \large 1. Introduction} 

\medskip 

Many aspects of the matrix formulation of $M$-theory \cite{000} has
been discussed in the past year. In this formulation, the compactified
$M$-theory is described by a non-abelian Yang-Mills theory living on
the dual torus \cite{030}. Different ways of compactification result
in a state degeneracy. For black holes this degeneracy is counted by
the Bekenstein-Hawking entropy, given by the area of the horizon. If 
the BPS bound is saturated, it has
been shown that the entropy corresponds to the minimum of the
moduli-dependent central charge \cite{010}. Especially, for $N=2$
black holes \cite{020} this approach has been very fruitful. On the
other hand, for Yang-Mills theories it is rather difficult to determine
the state degeneracy and it is the aim of this letter to address this
question. The central charge of a given Yang-Mills configuration
depends on the volumes of wrapped branes, which give the moduli of the
configuration. By extremizing this moduli-dependent central charge
we find the entropy (state degeneracy).  We will especially focus on
configurations that correspond to 5-d black holes and the 5-d
string. For finite $N$ the 5-d string is compact and their entropy
coincides with the Bekenstein-Hawking entropy of the 4-d black hole. In
the second part we discuss the microscopic interpretation of the
entropy and propose as Yang-Mills theory the worldvolume theory a
gauged KK-monopole.

\vspace{.8truecm}

\noindent
{\bf \large 2. Extremal central charges for matrix black holes} 

\medskip

The central charge (or the BPS mass) is a functions of the moduli and
the minimum gives the entropy. Let us motivate this procedure on the
supergravity side and consider the compactification from 11 to 5
dimensions.  Wrapping branes around cycles, one can think of the
moduli as the asymptotic volumes of these cycles.  As we will see
below, while extremizing the moduli one has to make sure, that one
only varies these cycles and keep all others fixed. Like the internal
cycles, also the moduli appear as dual pair, e.g.\ for a 6-d compact
space a wrapped 2-brane around a 2-cycle is dual to a wrapped 5-brane
around a 4-cycle and a 2-brane wrapped only over a circle is dual to
5-brane wrapped over a 5-cycle.

To be concrete let us consider the case of the 5-d black hole, which
can be obtained by wrapping three $M$-2-branes. The metric is given by
\be010
ds^2 = - {1 \over (H_1 H_2 H_3)^{2/3}} dt^2 + (H_1 H_2 H_3)^{1/3} 
\left( dr^2 + r^2 d\Omega_3 \right) \ .
\ee
with three harmonic functions $H_i = h_i + q_i/r^2$, where 
$q_i$ are the three electric charges and the $h's$  parameterize the
moduli space. We can always choose a coordinate system, which
asymptotically becomes the Minkowski space. Thus we can set
$h_1 h_2 h_3 =1$ and our solution is determined by only
two moduli. The mass of this BPS black hole coincides with the 
susy central charge
\be020
|Z|= \sum_{i=1}^3 M_i = (h_1 h_2 q_3 + h_2 h_3 q_1 + h_3 h_1 q_2)/3 =
\hat V^i q_i.
\ee
where $\hat V^i$ are the asymptotic volumes of the 2-cycles.  Note,
since the $h's$ are dimensionless also the $\hat V's$ have no
dimensions, i.e.\ we have divided out proper powers of the Planck
length $l_p$. This is also convenient, because we want to calculate the
entropy (= degeneracy of states) which should be dimensionless.  Next,
in extremizing this expression we have to vary the 2-volumes (or the
dual 4-volumes) and keep the other moduli fix. This means especially,
that we have to fix the total volume of the internal space
\be025
1 = \hat V^6 = \hat V^1 \hat V^2 \hat V^3 \ 
\ee
($\hat V^6$ becomes a modulus if we would wrap a brane around the
complete compact space, which is the case for an instantonic
wrapped 5-brane).

Now, the entropy can be obtained by extremizing the central charge 
with respect to  $\hat V_i$ \cite{010}, i.e.
\be030
{\cal S}_{bh} \sim |Z|^{3/2}_{min} \qquad \mbox{with}: \qquad 
\left({\partial \over \partial \hat V_i} |Z|\right)_{min} =0 \ .
\ee
After replacing e.g.\ of $\hat V^3 = 1/(\hat V^1 \hat V^2)$ in 
(\ref{020}) one finds
\be040
{\cal S}_{bh} \sim \sqrt{q_1 q_2 q_3 } \ .
\ee
which coincides with the Bekenstein-Hawking entropy obtained from the
area of the horizon. The analog procedure can also be done for the
dual configuration of the magnetic string, which is a bound state of
three 5-branes. In this case, because the mass density saturates the
BPS bound, one obtains the entropy per unit world volume (i.e.\  for
a fixed point on the world volume). 

As next step we are going to discuss the analog procedure for the
matrix model. We want to obtain the entropy not by translating of
known black hole results, but by extremizing the Yang-Mills central
charges. We do not need any information from the black hole or black
string -- the extremization yields the right result. Again, we want to
consider only special examples.

The $M$-theory, compactified on $T^d$ with the volume $V$, is
described  by a Yang-Mills theory living on the dual
torus with the volume and Yang Mills coupling constant
\be050
\Sigma = {l_p^{3d} \over R^d V} = {l_s^{2d} \over V}
\qquad , \qquad
g^2 = {l_p^{3(d-2)} \over R^{d-3} V} = {R^3 \Sigma \over l_p^6}
\ee
with $l_p$ as Plank length and $R$ is the radius of the $11th$ direction. 
Again we want to use dimensionless quantities and define
\be060
\hat g^2 = {g^2 \over l_s^{d-3}} = \left({l_s \over l_p}\right)^{d-3} 
{1 \over \hat V_d} \qquad , \qquad 
\hat \Sigma_d =  {\Sigma_d \over l_s^d} = \left({l_s \over l_p}\right)^d 
{1 \over \hat V_d} \
\ee
where $\hat V_d = V_d /l_p^d$ ; $l_s^2 = l_p^3/R$.
Then the central charges can be written as integrals over the 
dual Yang-Mills $d$-torus
\be070
\ba{lr}
\hat Z^{12} = {1 \over \hat g^2}  \int \hat \omega \wedge F  =
{1\over \hat g^2} \hat \Sigma^{d-2} \, m_{12} & 
\mbox{(transversal\ 2-brane)} \\
\hat Z^{1234} = {1 \over \hat g^2}  \int \hat \omega \wedge F \wedge F =
{1 \over \hat g^2} \hat \Sigma^{d-4} \, k_{1234} 
& \mbox{(wrapped\ 5-brane)} \\
\hat Z^{0} = {1 \over \hat g^2}  \int \hat \omega = {\hat \Sigma^d \over
\hat g^2} \, N & \mbox{(0-branes)} \\
\hat Z^{i} = {1 \over \hat g^2} \int \hat \omega \wedge e  =
{1\over \hat g^2} \hat \Sigma^{d-1} \, p_i & \mbox{(longitudinal\ 2-brane)} 
\ea
\ee
where we introduced $\int \hat \omega_d = \hat \Sigma^d$, the flux
number $m_{12}$ ($\int_{T^{12}} F = m_{12}$), the instanton number
$k_{1234}$ ($\int_{T^{1234}} F\wedge F = k_{1234}$) and the momentum
$p_i$ ($\int_{T^i} e = \int F_{0j} F_{ij} + .. = p_i$) (where the
integrals include the traces, see also \cite{050}). These central
charges coincide with the expressions of the sugra
side. Consider, e.g., the transversal 5-brane and using (\ref{060})
we find
\be080
\hat Z^{1234} = \left({l_s \over l_p}\right)  \left({R \over l_p}\right) 
\, \left({V^{1234} \over l_p^4}\right) k_{1234} = 
\left({l_s \over l_p}\right) \hat R \, \hat  V^{1234} k_{1234} \ .
\ee
This is the known central charge (or mass) contribution of a
transversal 5-brane, up to the prefactor $l_s/l_p$ which 
is a consequence of using different parameters to get dimensionless 
central charges. Introducing again the mass dimensions, both
central charges are identical
\be090
Z_{YM} = {\hat Z_{YM} \over l_s} = {\hat Z_{grav} \over l_p} = Z_{grav} \ .
\ee
The same procedure can be repeated for the other central charges in
(\ref{070}).

Analog to the black holes we will now build bound states and extremize
the total central charge. We will start with the analog configuration
to the 5-d black hole (\ref{010}), which is a threshold bound state of
three 2-branes.  Adding up the contributions, the total central charge
is given by
\be100
\hat Z = \sum_i \hat Z^i =  {1 \over \hat g^2} \, 
\int_{V_{\Sigma}} \omega \wedge F = {1 \over \hat g^2} (  
\hat \Sigma^1 \hat \Sigma^2 m_3 + \hat \Sigma^2 \hat \Sigma^3 m_1 + 
\hat \Sigma^3 \hat \Sigma^1 m_2)
\ee
with $i=1..3$ counting the different fluxes $(m_1, m_2 , m_3) \equiv (m_{12} ,
m_{34} , m_{56})$ through the different 2-tori $\Sigma^i$.

As for the black holes we have to vary only the moduli related to this
configuration and keep all others fix, i.e.\ we vary only 2- or
4-cycle volumes. Especially we have to keep fix the total volume of
the Yang-Mills torus and the Yang-Mills coupling constant
\be110
\hat \Sigma^1 \hat \Sigma^2 \hat \Sigma^3 = \hat \Sigma_6 = 1 \qquad ,
\qquad \hat g =1 \  .  
\ee 
Using this relation the minimum of (\ref{100}) is given
\be120 
{\cal S}^{YM}_{bh} \sim |\hat Z|^{3/2}_{min} = \sqrt{m_{12} m_{34} m_{56}} \ .
\ee
which coincides with the black hole entropy ({\ref{040})
(flux numbers in the matrix model correspond to the membrane
charges).

As second example we consider the pure 5-brane configuration ($5
\times 5 \times 5$), where the 5-branes are wrapped around 4-cycles 
which pairwise overlap on 2-cycles. This threshold bound state has 
the total central charge 
($(\hat Z^1 , \hat Z^2 , \hat Z^3) \equiv (\hat Z^{1234} , 
\hat Z^{1256} , \hat Z^{3456})$)
\be130
\hat Z = \sum_i \hat Z^i =  {1 \over \hat g^2} \, 
\int_{V_{\Sigma}} \omega \wedge F \wedge F = {1 \over \hat g^2} (  
\hat \Sigma^1 k_1 + \hat \Sigma^2 k_2 + \hat \Sigma^3 k_3 )
\ee
Again, in extremizing one has to vary only the 2-cycles, i.e.\ keeping
fix the 6-volume and the coupling constant like in (\ref{110}) and one
obtains
\be140
{\cal S}^{YM}_{str} \sim |\hat Z |^2_{min} = 
\left( k_1 k_2 k_3 \right)^{2/3} \ .
\ee
which is the entropy density of the magnetic string. In comparison to
the black hole entropy (\ref{120}) the different power here can be
understood from the different dimensionality of the horizons, namely
the 5-d black hole has an $S_3$ horizon whereas the string an $S_2$.
Note, for an infinite extended string it does not make sense to talk
about the total horizon -- for any fixed point of the world volume the
horizon is an $S_2$.

The cases so far correspond to the infinite momentum frame, but what about
the finite $N$ case \cite{110}? This means for the string that the radius is
finite and effectively describes the 4-d black hole case.
In this case we have to add 0-brane contributions to the central
charge in (\ref{130}) and get
\be150
\hat Z = {1 \over \hat g^2}(\hat \Sigma^6 N +  \hat \Sigma^1 k_1 + 
\hat \Sigma^2 k_2 + \hat \Sigma^3 k_3 ) \ .
\ee
This additional contribution also imply that we have
to take into account a further modulus. On the sugra side the
additional modulus is the radius or better the dimensionless quantity $\hat R
= R/l_p$. In analogy we can take on the Yang Mills side the
dimensionless string length $l_s/l_p$. But $l_s$ is implicitly
contained in all $\Sigma$'s and in $\hat g$, because we used $l_s$ to
make these quantities dimensionless, e.g.\ by $\Sigma^i \rightarrow
\Sigma^i / l_s^2 = \hat \Sigma^i$, see (\ref{060}).  Therefore, we
should replace $l_s$, which can be done by 
$\hat \Sigma^i \rightarrow l_s^2 / l_p^2\,
\hat \Sigma^i$.  Doing this everywhere, also for the gauge coupling,
we obtain
\be160
(l_p/l_s ) \, \hat Z = { 1 \over (l_s/l_p)^3 \hat g^2}
 ((l_s/l_p)^6 \Sigma^6 N +  (l_s/l_p)^2( \hat \Sigma^1  k_1 + 
\hat \Sigma^2 k_2 + \hat \Sigma^3 k_3 )) 
\ee
(note $\hat Z$ has just the inverse prefactor, because it was obtain
by $\hat Z = l_s Z$) and be written as
\be170
\hat Z = { 1 \over \hat g^2}
 \left((l_s/l_p)^4 \, \Sigma^6 N +  ( \hat \Sigma^1  k_1 + 
 \hat \Sigma^2 k_2 + \hat \Sigma^3 k_3 ) \right) \ .
\ee
The proper constraint is then given by
\be181
(l_s/l_p)^2 \, \hat \Sigma^1 \hat \Sigma^2 \hat \Sigma^3 =1 \qquad , \quad
\hat g =1
\ee
which is completely analog to the sugra constraint, where the 7-volume
has to be fixed
\be191
\hat R \hat V^1 \hat V^2 \hat V^3 = (l_p/l_s)^2 \hat V^1 \hat V^2 \hat V^3
=1 \ .
\ee
Note, in terms of (\ref{060}) we could use also the $\hat V$'s as moduli. 
Furthermore, instead of taking $l_s$ as additional moduli one could also 
take $g_0$ (the coupling constant of the 0+1 quantum mechanics) and
$g_s$ on the sugra side. But in any case, one has to fix the
gauge coupling $\hat g$ resp.\ the 4-d Newton constant on the sugra side.

Using this constraint and extremizing with respect to $\hat \Sigma^i$ 
gives for the entropy
\be195
{\cal S}^{YM}_{4d-bh} \sim |\hat Z |^2_{min}= \sqrt{N k_1 k_2 k_3}\ 
\ee
which again coincides entropy of 4-d black holes.

The same procedure can be used also for other configurations.
Considering e.g.\ the configuration of $2 \times 5 + mom$ and 
compactify it on a $T^5$ we obtain for the central charge
\be101
\hat Z = \hat Z^0 + \hat Z^{1} + \hat Z^{2345} 
\ee
and after inserting the expressions and performing the minimization
procedure we find
\be135
{\cal S}^{YM}_{bh} \sim |Z|^{3/2}_{min} =  \sqrt{N p \, k_{1234}} \ .
\ee
which is the entropy of the 5-d black hole.

Note, that this procedure is not specific to torus compactification.
It should be applicable to much more general cases like $K3$
\cite{180} or even Calabi-Yau compactification.  What one only needs
are non-trivial 2- and 4-cycles giving magnetic fluxes and instanton
numbers and their radii giving the moduli.


\vspace{.8truecm}

\noindent
{\bf \large 3. The Yang Mills theory and the microscopic picture} 

\medskip 

In order to discuss the microscopic interpretation of the entropy (=
minimum of the central charge) we have to consider the Yang Mills
theory. So far we assumed that a Yang Mills description exists.

As long as one compactifies the $M$-theory up to a 3-torus, the Yang
Mills theory is well defined. However for $d>3$, one has to address
the non-renormalizibility of the ``standard YM theory''. In recent
times, one has tried to overcome this problem by considering
worldvolume theories of branes, which decouple from the bulk in a
certain limit.  Hence, they can serve as theories describing
compactification for $d>3$.  This has been done for $d \le 5$ by
taking the worldvolume theory of the $NS$-5-brane \cite{120}. In order
to obtain a $d=6$ compactification one has discussed the worldvolume
of the $KK$ monopole in 11-d \cite{080}, which has a gauge field
enhancement at points where 2-cycles of the Taub-NUT space collapse,
see e.g.\ \cite{130}. Since this theory contains again membranes, it
has been suggested to formulate this theory in terms of a matrix model
as well \cite{090}. However, in recent discussion \cite{070} it has
been argued that for the ``standard KK-brane'' it is difficult to see
how the decoupling can go. The reason is, that in the expected
decoupling limit the brane effectively disappears leaving a space
with $A_{N-1}$ singularities. This can also be described by 
interaction with graviton modes (0-branes in 10-d) in the compact 
KK-direction. Hence, one has to find a way to decouple these modes
from the brane world volume.

As discussed in \cite{170} the 11-d worldvolume theory of the KK
monopole can be seen as a pure gravitational brane ($G$-brane) and
since the circular isometry has no natural worldvolume interpretation,
this is a 6-brane. There is however a subtlety with this brane. Naively
one would expect, that a 6-brane in 11-d gives raise to 4 scalars (the
4 transversal directions). However, taking into account also the
degrees of freedom of the worldvolume vector, this does not fit in
known 7-d supersymmetric theories. Hence, one has to eliminate one
degree of freedom. As suggested in \cite{070}, this can be done by
gauging the circular isometry of the monopole and one obtains as
Born-Infeld action
\be180
 S_{KK} \sim \int d^7 \xi \, k^2 \sqrt{| 
 \det\left[ \partial_i X^{\mu} \partial_j X^{\nu} \Pi_{\mu\nu}  + k^{-1} 
 (F_{ij} - k^{\mu} \, \partial_i X^{\nu} \partial_j X^{\rho} \,
 C_{\mu\nu\rho})\right]|} + S_{WZ} \ , 
\ee
where $\Pi_{\mu\nu} = g_{\mu\nu} - k^{-2} k_{\mu} k_{\nu}$, $k^2= k^{\mu}
k^{\nu} g_{\mu\nu}$ and $g_{\mu\nu}$ is the usual 11-d KK monopole solution
($M_7 \times Taub-NUT$); $F_{ij}$ is the world volume gauge field and
$k^{\mu}$ the Killing vector related to the isometry that has been
gauged.

After this gauging the coordinate in the isometry direction ($X^{\mu}=
k^{\mu}$) decouples from the brane, because $\Pi_{\mu\nu}$ is a
projector on the space transversal to the Killing vector and the
3-form potential $C_{\mu\nu\rho}$ is contracted with
$k^{\mu}$. Therefore, by gauging, one (isometry) direction has been
hidden and one introduced a new parameter $k^{\mu}$, which scales with
the radius of the ``hidden direction''.  Expanding the above action
one realizes that $k$ does not enter the gauge field coupling $1/g^2 =
M_{Pl}^3$. On the other hand $k$ acts as a coupling constant, like the
dilaton in $D=10$. Thus, taking the limit $k \to \infty$ keeps the
worldvolume gauge theory, but suppresses all interactions, especially
it decouples the bulk theory. In order to make this statement more
explicit, one has to couple the above action to 11-d sugra, e.g.\ by
considering the theory $S = S_{11} + S_{KK}$ and investigates the
equations of motion (see \cite{070}). In any case sending $k
\rightarrow \infty$ one has suppressed all gravity.  This procedure is
very similar to the $NS$-5-brane worldvolume theory discussed in
\cite{120}, which is a string theory that decouples from the 10-d bulk
theory in the limit of vanishing string coupling and finite string
mass. Similar here, the field theory (\ref{180}) decouples from the
bulk in the limit of vanishing membrane coupling $1/k$ and finite
Planck mass.

For this decoupling it is essential, that one considers the 11-d KK
action, i.e.\ (\ref{180}) contains the 11-d metric and
$C$-field. Compactifying this theory yields in 10-d the known 6-brane
solution with $k$ giving the dilaton. But since one has to rescale the
metric in going from 11 to 10 dimensions, the 10-d Yang-Mills coupling
is dilaton dependent! Note, coming from the 10-d 6-brane one has to
decompactify it in a non-trivial way, where the $11th$ direction
decouple from the brane (and also corresponding gravitons moving along 
this direction). 

Now we can start with the discussion of the microscopic picture
yielding the Yang-Mills entropy discussed in the last section. We will
not identify all states neither we discuss the complete $U$-duality
group.  Instead, our aim is to give arguments why the entropy formulae
are related to the degeneracy of states. For the black holes / black
strings this has been done using $D$-brane techniques, but for the
Yang-Mills formulation our understanding is still not yet complete.

The $M$-theory compactification of $5 \times 5 \times 5$ is described by
a bound states of instantons only. For this configuration the Wess-Zumino
part, that has to be added to the Born-Infeld action, contains
\cite{040}
\be190
S_{WZ} = \int C_3 \wedge Tr( F \wedge F)
\ee
where $C_3$ is the 3-form potential.
Since a 5-brane corresponds to a non-trivial instanton configuration,
it is a source for a (instantonic) membrane on the world volume
theory.  However this brane is not the usual localized brane, instead
it is smeared over a certain region of space time which is mainly
given by the instanton size. In the limit of vanishing instanton size
the source becomes singular and represents a ``standard'' membrane
lying inside a 6-brane. This is in complete analogy to strings
appearing in a 5-brane for vanishing instanton size \cite{150}.

It is now tempting to do the microscopic state counting in terms of
these brane states, e.g.\ the instanton number translates into the
charge of the membrane or equivalently the number of parallel
membranes. Following this procedure, the configuration yielding the
entropy ($5\times 5 \times 5 + mom$) corresponds in the
Yang-Mills picture for vanishing instanton size to a configuration of
three membranes intersecting over points. The state degeneracy of the
instanton bound state should coincides with the state degeneracy of
the membrane bound state, which counts the possibilities of wrapping
three membranes around 2-cycles of the 6-torus. In the Yang Mills
picture the longitudinal momentum modes counted by $N$ give the rang
of the $U(N)$ group. In the analog brane picture for small
instantons, it is suggestive to see these modes as 6-branes wrapping
the complete $T^6$. By this procedure, one can reduce the entropy
counting to a brane counting using the $D$-brane technique.

Let us discuss this procedure more explicit for the $M$-theory
configuration $2\times 5 + mom$ with the $11th$ direction lying along
the common worldvolume of the 2- and 5-brane (see also \cite{050},
\cite{060}).  In the infinite momentum frame this case correspond to
the 6-d dyonic string and for finite $N$ the string becomes compact
and gives the 5-d black hole. In this case we have to consider a 5+1
dimensional Yang-Mills theory. The translation is now as follows: the
(sugra) 5-brane corresponds to an instantonic strings and the 2-brane
to momentum modes traveling along this string and the momentum is
translated to the number of (YM) 5-branes.  Therefore, the state
counting is reduced to a counting of momentum modes for a
string. Again we can argue, that for shrinking instanton size we
obtain the known brane configuration of a string lying inside a
5-brane. The statistical entropy that counts string states is
$S_{stat.} = 2 \pi \sqrt{c p/6}$, where $c=3/2 \, D_{eff}$ and
$D_{eff}$ is the effective dimension where one can distribute the
momentum modes counted by $p$.  In our case $D_{eff}$ is given by the
dimension of the moduli space of $k$ strings (=number of instantons)
inside of $N$ 5-branes, which is given by \cite{040} $D_{eff} = 4
k(N+1) $. As result the statistical entropy is given by $S_{stat} = 2
\pi \sqrt{k(N+1) p}$ and coincides for large $N$ with (\ref{135}).

In conclusion, the aim of this letter was to discuss the state
degeneracy (or entropy) in matrix models. 
Following analog procedures from supersymmetric black holes,
we argued that the entropy is given by the minimum of the
moduli-dependent central charge.  In the second part we discussed a
way to count the microstates for the Yang-Mills configuration. The
main tool was to employ the fact, that the Yang-Mills fields act as
sources for new branes. The degeneracy of these brane configuration
can be counted by using the $D$-brane technique. Note, we were
counting Yang-Mills states and not the states of the supergravity
side! For this counting we only used the facts: (i) that the
Yang-Mills configuration is equivalent to intersecting brane
configuration for vanishing instanton size and (ii) that the
degeneracy of states should not alter if we shrink the instanton
size. As 7-d Yang-Mills theory that is needed for the $T^6$
compactification, we discussed the world volume theory of a gauged
11-d KK-monopole. The gauging effectively introduces a membrane
coupling constant and in the limit of vanishing coupling the 6+1
dimensional field theory decouples from the bulk.


\vspace{1truecm}

{\bf Acknowledgments}

I would like to thank especially Joachim Rahmfeld for numerous
discussions and for collaborating in early stages of this work.  In
addition I am grateful Eric Bergshoeff for valuable discussions and
the Stanford University for its hospitality, where this work has been
done.  The work is supported by the Deutsche Forschungsgemeinschaft
(DFG).



\end{document}